\documentclass[runningheads]{llncs}
\usepackage{amsmath}
\usepackage{graphicx}
\usepackage{comment}
\usepackage{booktabs}
\usepackage{multirow}
\usepackage{float}
\PassOptionsToPackage{hyphens}{url}
\usepackage{hyperref}
\restylefloat{table}
\usepackage{algorithm}
\usepackage{algpseudocode}
\usepackage{xcolor,soul}

\floatname{algorithm}{Procedure}

\DeclareRobustCommand{\hlgray}[1]{{\sethlcolor{lightgray}\hl{#1}}}

\begin{document}

\title{A Cost-Sensitive Meta-Learning Strategy for Fair Provider Exposure in Recommendation}
\titlerunning{Cost-Sensitive Meta-Learning Strategy for Fair Provider Exposure}

\author{%
Ludovico Boratto\inst{}\orcidID{0000-0002-6053-3015} 
\and
Giulia Cerniglia\inst{}
\and
Mirko Marras\thanks{Corresponding author.}\inst{}\orcidID{0000-0003-1989-6057} 
\and
Alessandra Perniciano\inst{}\orcidID{0009-0003-8956-5058} 
\and
Barbara Pes\inst{}\orcidID{0000-0003-3983-6844}
}

\authorrunning{Boratto et al.}

\institute{University of Cagliari, Cagliari, Italy\\
\email{\{ludovico.boratto,mirko.marras\}@acm.org, g.cerniglia1@studenti.unica.it, \{alessandra.pernician,pes\}@unica.it}}

\maketitle        

\begin{abstract}
When devising recommendation services, it is important to account for the interests of all content providers, encompassing not only newcomers but also minority demographic groups. In various instances, certain provider groups find themselves underrepresented in the item catalog, a situation that can influence recommendation results. Hence, platform owners often seek to regulate the exposure of these provider groups in the recommended lists. In this paper, we propose a novel cost-sensitive approach designed to guarantee these target exposure levels in pairwise recommendation models. This approach quantifies, and consequently mitigate, the discrepancies between the volume of recommendations allocated to groups and their contribution in the item catalog, under the principle of equity. Our results show that this approach, while aligning groups exposure with their assigned levels, does not compromise to the original recommendation utility. Source code and pre-processed data can be retrieved at \url{https://github.com/alessandraperniciano/meta-learning-strategy-fair-provider-exposure}.
\end{abstract}

\section{Introduction} 
Online platforms serve as intermediaries that connect various parties, typically consumers and providers. These platforms frequently rely on recommender systems that employ predictive relevance to match consumers with suitable items offered by providers \cite{DBLP:reference/sp/2022rsh}. Traditionally, optimizing recommendations solely for the benefit of consumers has been the ultimate objective. However, it becomes important to recognize that recommender systems exist within multi-sided environments and, as such, should carefully consider interests and needs of providers, going beyond the scope of consumer-centric concerns only \cite{DBLP:journals/umuai/AbdollahpouriAB20}.

Notable concerns pertain to the unfair exposure of provider groups in the recommended lists, which goes beyond just legal aspects. It can also be analyzed from the perspective of a platform's business model \cite{muzellec2015two}. For example, platforms like Coursera and GoFundMe may prioritize ensuring that new instructors or projects receive adequate exposure in the recommendations, even though during the initial period of inclusion in the item catalog, they naturally engaged fewer users than established providers \cite{DBLP:conf/ecir/GomezBS21}. Mitigating these exposure discrepancies can not only counter a possible discrimination but also bring beyond-utility benefits, such as introducing novelty and diversity \cite{DBLP:journals/tiis/KaminskasB17}. Consequently, supporting provider groups, especially those belonging to minorities, becomes imperative \cite{DBLP:journals/umuai/BorattoFM21,DBLP:journals/ipm/GomezBS22,DBLP:conf/sigir/GomezZBSM21}.

Several mitigation methods have been proposed to reorganize the initial lists of a recommender system, aiming to meet specific objectives related to diversification. These methods frequently concentrate solely on the aspect of visibility \cite{DBLP:conf/ssdbm/YangS17,DBLP:conf/cikm/ZehlikeB0HMB17}. Visibility, in this context, refers to the fraction of items from a particular provider group in the uppermost segment of the rankings. Conversely, there are other approaches, often introduced within a fairness framework, which strive to guarantee an equitable exposure among provider groups \cite{DBLP:conf/sigir/BiegaGW18,DBLP:conf/ecir/GomezBS21,DBLP:conf/kdd/SinghJ18}. However, these methods do not consistently address situations in which items from minority groups end up being ranked at the lower end of the ranking segment.

Our approach in this study stands out in four key ways. First, it addresses position bias in ranked results, which is overlooked in diversification-related fair ranking research \cite{DBLP:conf/ssdbm/YangS17,DBLP:conf/cikm/ZehlikeB0HMB17}. Second, in contrast to \cite{DBLP:conf/sigir/BiegaGW18}, it does not depend on metrics that require a rank-aware quality assessment, making it more practical for recommender systems. Third, it achieves balancing objectives for user rankings without the need for extensive feedback data from all users, in contrast to \cite{DBLP:conf/ecir/GomezBS21}. Finally, \cite{DBLP:conf/sigir/MarrasBRF22} can control provider group exposure, but its complexity is directly tied to the number of users and items, posing challenges when applied to large datasets commonly encountered in real-world applications. The computational complexity of our approach in this paper does not depend on these factors.

In this paper, we examine a scenario where providers are categorized based on a shared characteristic, and certain provider groups have limited representation of items in the catalog and, subsequently, in the recommendations. As a first novel contribution, we introduce a cost-sensitive meta-learning method to ensure that the level of exposure for provider groups resembles their contribution in the catalog, guided by the principle of equity \cite{DBLP:conf/sigir/BiegaGW18}. As a second novel contribution, we conduct a comprehensive evaluation and discuss our approach with respect to state-of-the-art alternatives, to assess how supporting provider groups influences aspects of utility and beyond-utility on two distinct public data sets.

\section{Methodology}

\subsection{Fairness Objective Definition}
Our study centers on the exploration of a recommendation policy having a primary objective: securing an equitable distribution of exposure for provider groups relative to their representation within the catalog. This policy's philosophy aligns with the well-known distributive norm based on equity \cite{DBLP:conf/sigir/BiegaGW18,DBLP:conf/iccsa/FenuLM20,DBLP:conf/interspeech/FenuMMM21}. To illustrate the importance of this policy, consider an online course platform that hosts some courses provided by male instructors and others sourced from female instructors\footnote{While gender is by no means a binary construct, to the best of our knowledge, no data set with non-binary gender exists. What we are considering is a binary feature.}. With no intervention, the platform may inadvertently favor courses from one group, possibly reinforcing disparities. Following our policy, courses provided by male and female instructors receive proportionate exposure in recommendations, leading to platform that empowers instructors irrespective of gender and reduces any concentration of attention on a given group.

\begin{table}[!t]
\caption{\footnotesize In addition to the number of users, we show the number of items in the catalog and the percentage of those provided by females. Then, we place the total number of interactions and of their subset in the train set, and the percentage of train interactions for items from female providers. We also show the percentage of train triplets with a positive/negative item from a female provider, under a random sampling.}.
\label{tab:statistics_datasets}
    \resizebox{1.0\linewidth}{!}{
\begin{tabular}{r|r|rr|rrr|rr}
\multirow{2}{*}{\textbf{Data Set}} & \multirow{2}{*}{\textbf{\# Users}} & \multicolumn{2}{c|}{\textbf{Items}}                      & \multicolumn{3}{c|}{\textbf{Interactions}}                                                           & \multicolumn{2}{c}{\textbf{\% Female in Triplets}}                 \\
                                   &                                    & \multicolumn{1}{r|}{\textbf{Total}} & \textbf{\% Female} & \multicolumn{1}{r|}{\textbf{Total}} & \multicolumn{1}{r|}{\textbf{Train}} & \textbf{\% Female} & \multicolumn{1}{r|}{\textbf{Pos. Item}} & \textbf{Neg. Item} \\ \hline
ML-1M                              & 6,040                              & \multicolumn{1}{r|}{3,876}          & 6.30 \%            & \multicolumn{1}{r|}{998,539}        & \multicolumn{1}{r|}{634,640}        & 3.60 \%                  & \multicolumn{1}{r|}{3.40 \%}               & 6.40 \%               \\ \hline
COCO                               & 73,022                             & \multicolumn{1}{r|}{34,442}         & 12.70 \%           & \multicolumn{1}{r|}{520,531}        & \multicolumn{1}{r|}{294,271}        & 7.60 \%                  & \multicolumn{1}{r|}{7.50 \%}               & 12.70 \%              \\ \hline
\end{tabular}}
\vspace{-2mm}
\end{table}

\subsection{Data Preparation}
To align with benchmarks in prior work \cite{DBLP:journals/umuai/BorattoFM21,DBLP:conf/sigir/MarrasBRF22}, we consider experimenting with the target policy through our approach in two scenarios, namely movie recommendation and online course recommendation. To this end, we used two publicly available data sets, whose main statistics are described in Table \ref{tab:statistics_datasets}. They are notable for the inclusion of providers' gender labels, a rarity in such collections.

\textbf{MovieLens-1M (ML-1M)} \cite{harper2015} encompasses ratings for movies collected from users. Movie directors are traditionally assumed to be the providers in this scenario. We considered the gender labels from \cite{DBLP:journals/umuai/BorattoFM21}. As the main preprocessing steps, we (i) retrieved the directors and their gender from public sources for those movies this information was unknown, (ii) kept only the first director for movies with multiple directors primarily because s/he is typically the most influential in the film-making process (and to prevent duplication of directors in different sets), and (iii) filtered interactions such that items had at least 10 interactions since extreme cold-start scenarios are not our main focus.

\textbf{COCO} \cite{dessi2018} consists of ratings assigned to courses by learners, collected from an online course platform. For this benchmark, following prior work, we assumed that the course instructors represent the providers \cite{DBLP:conf/sigir/GomezZBSM21}. Similar to the preprocessing steps for ML-1M, to address the issue of missing gender information, we removed instructors and their associated courses, for those instructors we could not retrieve their gender from public sources. Then, to ensure enough input for personalization and avoid extreme cold-start scenarios, we filtered the data such that each course had at least 5 ratings and each learner gave at least 5 ratings. 

On both the considered data sets, we performed a temporal train-test split, allocating the most recent 20\% of interactions to the test set, another 20\% to the validation set, and the remaining to the train set.

\subsection{Pairwise Recommendation Model Creation}
Once the two considered datasets were pre-processed, we leveraged them to train a series of recommendation models. In our study, we focused on models optimized through pairwise learning, which forms the basis of several state-of-the-art personalized ranking algorithms \cite{zhang2019} and is widely studied in the context of methods pursuing fairness or diversification objectives, e.g., in \cite{DBLP:journals/umuai/BorattoFM21,DBLP:conf/sigir/MarrasBRF22}. Specifically, we employed the Bayesian-Personalized Ranking with Matrix Factorization (BPRMF) method, structured around an optimization criterion derived from a Bayesian analysis. It uses the prior probability for the model and a likelihood function that calculates the probability of a user preferring an item $i$ the user previously interacted with over another item $j$ the user did not interacted with. We adopted the BPRMF implementation provided in Elliot for better reproducibility \cite{anelli2021}. 

\subsection{Traditional Strategy: Pairwise Training via Random Sampling} 
To conduct an initial assessment about fairness among provider groups under the equity-based policy, we trained a recommendation model based on BPRMF, by following the traditional training protocol adopted in \cite{zhang2019}, as an example. To this end, we used a bootstrap sampling with replacement to sample train triplets for a user. Each sample consists of a triplet $(u, i, j)$, where 'u' is the user's index, 'i' is the positive item (i.e., the user interacted with), and 'j' is the negative item (i.e., the user did not interact with). These triplets are created by randomly sampling items according to the interactions in the train set. To initialize the 10-sized latent factor matrices for both users and items within the BPRMF's model architecture, we assigned values uniformly distributed between 0 and 1. The training process ran for 10 epochs, and for optimization, we utilized SGD with a learning rate of 0.001. Detailed (hyper-)parameter values are listed in the source code repository for a better reproducibility. 

\subsection{Our Proposal: Pairwise Training via Cost-Sensitive Sampling}

\textbf{Motivation}. Once the baseline recommendation models were trained on both data sets, we performed an assessment of fairness aimed to uncover potential discrepancies between the representation of different provider groups in the item catalog, the train interactions (and consequently the train triplets), and the provided recommendations. For convenience, let us assume to focus on the model trained on the ML-1M data set. Considering the top-10 recommended items, we observed that only $1.40\%$ of the exposure is given to items coming from a female director. This exposure level goes in contrast with the representation of female providers in the train interactions (3.60\%) and in the item catalog (6.30\%) - see Table \ref{tab:statistics_datasets}. We conjecture that this disparity in representation is rooted in the characteristics of the triplets sampled during training. While the percentage of triplets that involve a positive item from a female provider (3.40\%) is close to the representation of female providers in the train interactions (3.60\%), the same observation does not hold if we consider the negative items. The percentage of triplets that involve a negative item from a female provider (6.40\%) is disproportionately higher than their representation in the interactions and in the catalog. Hence, the model is fed with data that leads to disadvantage female providers more, since the model is optimized to predict a lower relevance for the negative item over the positive one for each triplet. 

\textbf{Intuition}. Guided from the abovementioned motivation, we propose to adopt a cost-sensitive learning strategy as a countermeasure. Cost-sensitive learning in data mining encompasses a set of algorithms tailored to account for different costs associated with specific characteristics of the problem to consider \cite{DBLP:books/sp/FernandezGGPKH18}. Under this class-cost perspective that, while training a model, more attention is given to training elements with higher associated costs (e.g., rare and valuable classes). While direct methods like CSTree \cite{DBLP:conf/icml/LingYWZ04} directly integrate costs into learning algorithms, meta-learning automatically transforms existing cost-insensitive models into cost-sensitive ones. It can be seen as a preliminary step that requires to pre-process train data from cost-insensitive learning methods (the BPR-MF model in our study), remaining applicable to any recommendation model. 

\begin{algorithm}[!b]
\caption{Cost-sensitive meta-learning sampling of train triplets}\label{alg:cap}
\begin{algorithmic}
\footnotesize
\Require $C \geq 1$ : Parametric cost for the majority group; $N \geq 0$: No. of train triplets.
\Ensure triplets : Generated triplets to train the pairwise recommendation model.
\State $p = [(C \times 100)/(C + 1), 100 / (C + 1)]$ 
\For{$i$ : \text{trainItems}}
\State probabilities(i) = isItemFromMajorityClass($i$) ? p[0] : p[1]
\EndFor
\State triplets = [\;]
\While{$N \geq 0 $} 
\State $u$ = sampleRandomly(listUsers)
\State $i$ = sampleRandomly(getUserItems($u$)))
\Repeat 
\State $j$ = sampleWithProb(trainItems, probabilities) 
\Until{listItems($u$).includes($j$)}
\State triplets.append(($u$, $i$, $j$))
\State $N = N-1$
\EndWhile
\end{algorithmic}
\end{algorithm}

\textbf{Proposal}. 
To validate our intuition, we therefore decided to propose and experiment with a novel parametric cost-sensitive meta-learning approach which adjusts the sampling distribution. This approach can regulate the exposure of provider groups in recommendations generated by the BPR-MF model and similar pairwise learning models, based on the contributions of these groups in the catalog. Procedure \ref{alg:cap} outlines the proposed sampling strategy for controlling the distribution of items from female and male providers within the negative item set of triplets. The parameter $C$ alters the distribution, allowing us to reduce or increase instances from the minority group in the negative or positive set and, consequently, influencing the relative exposure of different groups. The vector $p$ includes two elements: (i) the probability of selecting an item from the provider group deserving more emphasis (in the context of the negative set, this represents the majority group), and (ii) the probability of selecting the other group. In essence, by manipulating $C$, our approach provides control over the balance between female and male provider groups in the triplets, both in the positive and negative item sets. Our hypothesis is that, by aligning the proportions of items in the catalog from female and male providers in both positive and negative sets, the exposure of groups reflects the target fairness level. Note that our approach can be applied also for demographic attributes with more than two classes.    

\begin{table}[!b]
\vspace{-3mm}
\caption{\footnotesize Impact of the parameter $C$ on recommended lists (utility, exposure). For each data set, the gray row identifies the setting under which the representation of female items in the positive (Pos. Item) and negative (Neg. Item) item sets is the closest.}
\label{tab:results_impact_weights}
\centering
\resizebox{.95\linewidth}{!}{
\begin{tabular}{r|r|r||r|rr|rr}
\multirow{2}{*}{\textbf{Data Set}} & \multirow{2}{*}{\textbf{Item Type}} & \multirow{2}{*}{\textbf{Cost C}} & \multirow{2}{*}{\textbf{NDCG}} & \multicolumn{2}{r|}{\textbf{\% Female in Triplets}} & \multicolumn{2}{r}{\textbf{\% Female Exposure}} \\
                                   &                                     &                                  &                                & \textbf{Pos. Item}       & \textbf{Neg. Item}       & \textbf{Top 10}        & \textbf{Top 20}        \\ \hline
\multirow{4}{*}{ML-1M}             & -                                   & C = 1.0                     & 0.07                           & 3.4 \%                   & 6.4 \%                   & 1.4\%                  & 2.7\%                  \\
                                   & NEG                                 & C = 1.2                          & 0.07                           & 3.4 \%                   & 5.3 \%                   & 2.5\%                  & 2.8\%                  \\
                                   & \hlgray{NEG}                                 & \hlgray{C = 2.0}                            & \hlgray{0.07}                           & \hlgray{3.4 \%}                   & \hlgray{3.2 \%}                  & \hlgray{6.0\%}                  & \hlgray{3.9\%}                  \\
                                   & NEG                                 & C = 3.0                            & 0.07                           & 3.4 \%                   & 2.2 \%                   & 6.8\%                  & 4.5\%                  \\ \hline
\multirow{7}{*}{COCO}              & -                                   & C = 1.0                     & 0.02                           & 7.5 \%                   & 12.7 \%                  & 0.0\%                  & 0.0\%                  \\
                                   & NEG                                 & C = 2.0                            & 0.02                           & 7.5 \%                   & 6.6 \%                   & 0.0\%                  & 0.4\%                  \\
                                   & NEG                                 & C = 3.0                            & 0.02                           & 7.5 \%                   & 4.7 \%                   & 0.0\%                  & 2.2\%                  \\
                                   & NEG                                 & C = 5.0                            & 0.02                           & 7.5 \%                   & 2.9 \%                   & 0.0\%                  & 2.2\%                  \\
                                   & POS                                 & C = 1.2                          & 0.02                           & 10.5 \%                  & 12.7 \%                  & 2.1\%                  & 6.9\%                  \\
                                   & POS                                 & C = 2.0                            & 0.02                           & 10.9 \%                  & 12.7 \%                  & 5.3\%                  & 7.6\%                  \\
                                   & \hlgray{POS}                                & \hlgray{C = 3.0}                            & \hlgray{0.02}                           & \hlgray{12.3 \%}                  & \hlgray{12.7 \%}                  & \hlgray{12.2\%}                  & \hlgray{11.9\%}                  \\ \hline
\end{tabular}}
\vspace{-1mm}
\end{table}

\section{Results and Discussion} 
Given the intuitions we seek to validate, our analysis was focused on examining whether the proposed cost-sensitive meta-learning approach can lead to achieving the objective defined by the equity-based fairness policy and, if so, how it impacts recommendation utility. To this end, we compared the original pairwise model (Baseline, $C=1$), with our cost-sensitive approach under different cost settings. In Table \ref{tab:results_impact_weights}, we monitored recommendation utility (NDCG, with binary relevance scores, base-2 logarithm decay), representation in train data (\% Female Interactions, \% Female in Triplets), and group exposure in both the top-10 and top-20 recommended lists provided to users (\% Female Exposure).  

From the obtained results, we observed that our approach does not affect recommendation utility (NDCG column) in both data sets. In ML-1M, assigning a cost $C=2.0$ to the majority group (male items in this case) allowed us to have a very close representation of female items in both positive (3.4\%) and negative (3.2\%) elements of the triplets. As hypothesized, the exposure of items from female providers is also positively affected, reaching $6.0\%$. This is close to the representation of female providers in the catalog (6.4\% - see Table \ref{tab:statistics_datasets}). Differently from ML-1M, playing with the negative items during sampling did not allow to reach a similar representation of female items between positive and negative elements of the triplets in COCO. To this end, we applied the cost-sensitive approach to \emph{increase} the female representation in the \emph{positive} set. With this setting, under a cost $C=3.0$, we achieved such similar representation (12.3\% vs 12.7\%). Exposure was impacted, reaching $12.2\%$ on the top 10. Again, this is close to the representation of female providers in the catalog (12.7\% in Table \ref{tab:statistics_datasets}).  

In summary, by playing with the parameter $C$ to make the representation of the minority group equal in the positive and negative item sets right before the training phase, pairwise models trained on triplets sampled with our approach can lead to provider groups having an exposure estimate equal or substantially close to their representation in the catalog. No impact is observed in recommendation utility. Therefore, the proposed approach does not suffer from the typical trade-off between the overall recommendation utility of the recommender system and the fairness of exposure across provider groups in the recommended lists. Despite the importance of showing a more consolidated view of the existing methods for provider group fairness, any comparison to other state-of-the-art exposure control methods, such as LFRank \cite{DBLP:conf/ssdbm/YangS17}, FA*IR \cite{DBLP:conf/cikm/ZehlikeB0HMB17}, FOEIR \cite{DBLP:conf/kdd/SinghJ18}, GDE \cite{DBLP:conf/ecir/GomezBS21}, and MMR \cite{DBLP:conf/sigir/MarrasBRF22}, would not influence the overall assessment of the feasibility and effectiveness of our approach, which is the main focus of this paper. For instance, the method proposed in \cite{DBLP:conf/sigir/MarrasBRF22}, proven to lead to a better trade-off than all the others the original authors considered, still reports a non-negligible decrease in recommendation utility, between $5\%$ and $20\%$, under a comparable protocol.

\section{Conclusions and Future Work} 
In this paper, we introduced a cost-sensitive meta-learning approach to tackle the issue of imbalanced class distributions in train data for recommendation. By manipulating the sampling distribution, our method effectively manages provider group exposure in the recommended lists. The meta-learning nature of our approach makes it adaptable to different recommendation methods. Future work will involve applying our method to other pairwise recommendation methods and conducting extensive performance evaluations of effectiveness, fairness, and scalability across diverse domains. Moreover, for a better contextualization, we will perform an extensive comparison to fully assess the relative strengths and weaknesses of the proposed method against existing solutions under a wider range of metrics, including also the required effort to tune their (hyper-)parameters.

\vspace{2mm} \noindent \textbf{Acknowledgement}.
We acknowledge financial support under the National Recovery and Resilience Plan (NRRP), Miss. 4 Comp. 2 Inv. 1.5 - Call for tender No.3277 published on Dec 30, 2021 by the Italian Ministry of University and Research (MUR) funded by the European Union – NextGenerationEU. Prj. Code ECS0000038 eINS Ecosystem of Innovation for Next Generation Sardinia, CUP F53C22000430001, Grant Assignment Decree N. 1056, Jun 23, 2022 by the MUR.

\bibliographystyle{splncs04}
\bibliography{samplepaper}

\end{document}